# Coherent Control of ultracold molecular collisions: The role of resonances

Adrien Devolder[1], Timur V. Tscherbul[2], and Paul Brumer[1]*

[1]*Chemical Physics Theory Group, Department of Chemistry, and Center for Quantum Information and Quantum Control, University of Toronto, Toronto, Ontario, M5S 3H6, Canada*

[2]*Department of Physics, University of Nevada, Reno, NV, 89557, USA*

E-mail: adrien.devolder@utoronto.ca;ttscherbul@unr.edu,paul.brumer@utoronto.ca

**Abstract**

We consider the coherent control of ultracold molecule-molecule scattering, impacted by a dense set of rovibrational resonances. To characterize the resonance spectrum, a rudimentary model based on multichannel quantum defect theory has been used to study the control of the scattering cross section and the reaction rate. Complete control around resonance energies is shown to be possible, but thermal averaging over a large number of resonances significantly reduces the extent of control of reaction rates due to the random distribution of optimal control parameters between resonances. We show that measuring the extent of coherent control could be used to extract meaningful information about the relative contribution of direct scattering versus collision complex formation, as well as about the statistical regime.

# TOC Graphic

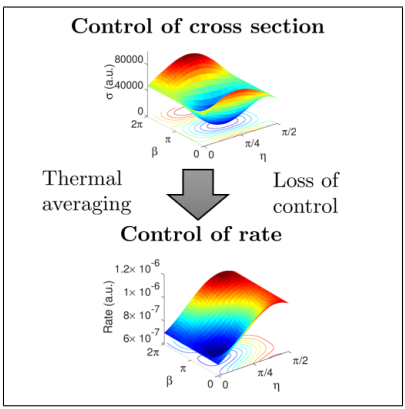

**Introduction.**– Ultracold dialkali molecule gases have opened up exciting opportunities in quantum information, many-body physics, precision measurement and controlled chemistry.[1–3] Unfortunately, undesirable trap losses from the confining trap, associated with molecular collisions, limit these opportunities. The collisional dynamics between two dialkali molecules involves a barrierless and deep potential energy surface, and is thought to be affected by a dense set of rovibrational resonances.[4,5] This proliferation of resonances is associated with the appearance of long-lived four-atom complexes,[6] which were experimentally observed for KRb-KRb scattering.[7] Two complex dissociation mechanisms can explain the observed trap losses. First, if the two species are reactive, the complex can dissociate into one of the reactive channels. Second, if the two species are non-reactive, then the complex can be photodissociated by the trapping light, as first theoretically proposed by Christianen *et al.*[8] and subsequently confirmed experimentally for RbCs.[9,10] However, the photodissociation mechanism does not explain the losses observed for NaRb and NaK molecules[11] and their explanation is still an open question. Furthermore, the proliferation of resonances relates to quantum chaos, as shown by many studies.[12–18]

The theoretical description of ultracold dialkali molecule-molecule scattering is complicated and, unfortunately, an accurate quantum scattering calculation is far beyond the computational state-of-the-art. However, certain aspects of the scattering can be described with statistical methods when there is an average over a large number of resonances. Previously used for nuclear reactions,[19,20] unimolecular reactions[21] and chemical reactions,[22,23] the statistical methods were first introduced into the field of ultracold molecular collisions by Mayle *et al.*[4] and have been widely used.[5,18,24–27] A key quantity of these methods is the density of states $\rho$, calculated by Mayle *et al.*[5] and subsequently improved by Christiaenen *et al.*[8] using a quasiclassical method. These calculated values of $\rho$ are too small to expect the applicability of statistical model, which requires a large value. However, the loss processes can broaden the resonances and the statistical regime can be reached due to the resonance overlap.[27] This may explain the agreement of some experiments with the statistical model.[28]

Another possible explanation is a higher value of $\rho$ thanks to symmetry breaking by external fields.[29] Understanding the applicability of the statistical model can be also investigated with atom-diatom collisions, simpler for numerical simulations. For example, the lifetime of the ultracold atom-molecule collision complex (Rb-KRb) was measured to be five orders of magnitude longer than predicted by RRKM theory,[30] suggesting a possible failure of the statistical description.

For potential applications, the control of trap losses is a crucial challenge. Different mechanisms for control have been proposed: electric shielding,[31–33] microwave shielding,[34–36] optical shielding[37] or magnetic control.[38,39] These methods rely upon on external fields, with molecules/atoms prepared in a well-defined internal state. However, the use of external fields could be undesirable for some applications. For example, the presence of an external field can induce large perturbations of molecular energy levels, limiting the precision of spectroscopic experiments. Significantly, coherent control, the focus of this article, is a field-free method, where the molecules are prepared in a superposition of internal states and the ultracold collisions are controlled by changing the coefficients of the initial superposition. The principle of coherent control lies in the resultant interference.[40] In previous work, we demonstrated that the ultracold regime is ideal for coherent control, showing complete suppression (or activation) of resonant processes such as collisional spin exchange[41] via destructive (constructive) interference. Furthermore, the total cross section can be extensively controlled via an initial superposition state, as we described by a generalization of the optical theorem.[42]

The absence of accurate coupled-channel calculations for ultracold dialkali molecule-molecule collisions is a challenge for the simulation of coherent control, which relies on the availability of full S-matrix elements including their phase. Otherwise, the precise prediction of the control landscape cannot be obtained. However, insight can be gained about the controllability as discussed below. Specifically, we introduce a rudimentary model which randomly generates samples of resonances. This allows us to study the control of two types of collisional quantities, defined with or without energy average over a large number of resonances, e.g. reaction rates and cross sections, respectively. The control of the cross section can be complete around isolated resonance energies while on the other hand, the overlap of many resonances limits the amount of control even at a single collisional energy. Similarly, the energy average over a large number of isolated resonances diminishes the extent of control of the reaction rate. These results also show the potential of coherent control as a probe of the statistical regime of ultracold scattering.

**Scattering model–** Consider a rudimentary model with $2+N_f$ open channels: 2 in the reactant arrangement (denoted $|1\rangle$ and $|2\rangle$) and $N_f$ in the product arrangement (denoted $|3\rangle, |4\rangle, ..., |N_f + 2\rangle$), for which the value has been fixed at 100 by default. The system is initially prepared in a superposition of channels $|1\rangle$ and $|2\rangle$, which corresponds to ultracold scattering for two different energetically degenerate two-molecule internal states in the s-wave regime:

$$|\Psi_{ini}\rangle = \cos\eta \, |1\rangle + \sin\eta \, e^{i\beta} \, |2\rangle . \quad (1)$$

Here $\eta = [0, \pi/2]$ and $\beta = [0, 2\pi]$ are the superposition parameters, related respectively to the relative population and phase between the different components $|1\rangle$ and $|2\rangle$ of the initial scattering state (1) .

To describe ultracold bimolecular scattering, we follow a Multichannel Quantum Defect Theory(MQDT) procedure,[4,26] which separates the short-range and the long-range dynamics. Details are given in the Supporting Information. First, the short-range reactance (K) matrix is parameterized by random generation of a sample of resonances with five parameters: $\rho$, $w$, $x$, $s_{dir}$ and $\Gamma_{loss}$. Here, $\rho$ is the density of resonances, $w$ is the Brody parameter which parametrizes the resonance energies, while $x$ is a parameter that determines the statistics of the resonance widths. $s_{dir}$ gives the extent of the direct scattering and finally, $\Gamma_{loss}$ is the loss rate to unobserved channels, e.g. after photoexcitation or collisions with a third molecule. By varying these parameters, different regimes can be reached. As the generation of the resonances in the model is random, the discussions of results focus on common properties observed for any sample of resonances. Here, we focus on the isolated resonance case (small $x$ and $\Gamma_{loss} = 0$) where the resonance widths $\Gamma_r^T$ are smaller than the mean level distance $d$, and on the lossy resonance case (small $x$ but large value for $\Gamma_{loss}$) which is the expected regime for dialkali-dialkali scattering.[27]

The long-range S-matrix is obtained from the short-range K-matrix via the usual MQDT algebraic procedure.[4,26] In our model, MQDT is only applied in the space of open channels while the closed channels, responsible for the resonances, are included in the parametrization of the short-range K-matrix.[4,5,26] This differs from other MQDT formulations which explicitly include the closed channels.[43] Using the S-matrix, the reactive cross section from the initial superposition (Eq. 1) to a product channel $|p\rangle$ is calculated as:

$$\sigma_{sup \to p}(\eta, \beta, E) = \frac{\pi}{k^2} \Big( \cos^2 \eta |S_{1p}|^2 + \sin^2 \eta |S_{2p}|^2 \\ + 2 \sin \eta \cos \eta \ Re(S_{1p} S_{2p}^* e^{-i\beta}) \Big). \quad (2)$$

where $k$ is the relative momenta and $E$ is the collisional energy.

The sum over all product channels then gives the total reaction cross section $\sigma_{sup \to reac} = \sum_{p=3}^{N_f+2} \sigma_{sup \to p}$, which provides the reaction rate as the thermal average:

$$k_{reac}(\eta, \beta) = \sqrt{\frac{8}{\pi \mu E_{th}}} \int \sigma_{sup \to reac}(E) \frac{E}{E_{th}} e^{-E/E_{th}} dE, \quad (3)$$

where $E_{th} = k_B T$ is the thermal energy, $k_B$ is the Boltzmann constant, $T = 500$ nK is the temperature of the ultracold molecular gas, and $\mu$ is the reduced mass of KRb-KRb. We note that even though calculations below are performed for KRb-KRb scattering, the main conclusions of the article are general and would apply to other dialkali-dialkali collisions. We quantify the extent of control with a control index defined as the ratio between the interference contribution and the geometric mean of the cross sections without superposition :

$$R_c(E) = \frac{|\sigma_{int}|}{\sqrt{\sigma_{1 \to reac} \sigma_{2 \to reac}}}. \quad (4)$$

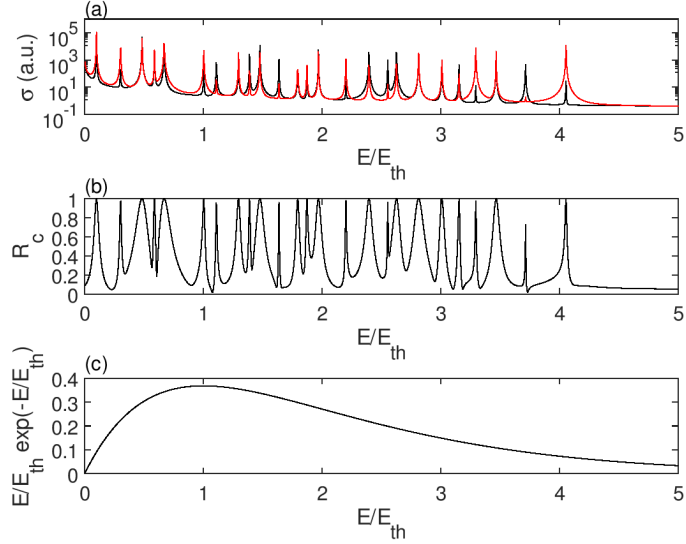

Figure 1: Energy dependence (a) of the cross section from an initial channel $|1\rangle$ ($\sigma_{1\to reac}$) (black) and $|2\rangle$ ($\sigma_{2\to reac}$) (red), and (b) of the control index ($R_c$) obtained from a representative resonance sample generated with $\rho = 5$, $x = 0.0001$, $s_{dir} = 0.0001$, $w = 1$ and $\Gamma = 0$. (c) Thermal distribution $\frac{E}{E_{th}}e^{-E/E_{th}}$

Here $\sigma_{int} = \frac{\pi}{k^2}\sum_{p=3}^{N_f+2} S_{1p}S_{2p}^*$ is the interference contribution to the cross section, and $\sigma_{1\to reac}$ and $\sigma_{2\to reac}$ are the reactive scattering cross section from the initial channel $|1\rangle$ and $|2\rangle$, respectively. Note that this definition of the control index relies on the Schwartz inequality, so that $R_c(E)$ lies between zero and one. A similar definition applies to a control index for the reaction rate:

$$R_{c,rate} = \frac{\left|\int \sigma_{int} E e^{-E/E_{th}} dE\right|}{\sqrt{\left(\int \sigma_{1\to reac} E e^{-E/E_{th}} dE\right)\left(\int \sigma_{2\to reac} E e^{-E/E_{th}} dE\right)}}. \qquad (5)$$

**Control of energy-resolved quantities (e.g. cross section).–** Here, we considers the case where the Maxwell-Boltzmann distribution of the molecular gas is sufficiently narrow such that individual scattering resonances are well resolved. First, consider the regime where the reaction is only driven by isolated resonances without substantial direct contribution, e.g $s_{dir} \ll 1$. We illustrate the control in this regime from a representative sample of resonances generated with the following parameters $\rho = 5$, $x = 0.0001$, $s_{dir} = 0.0001$, $w = 1$ and $\Gamma_{loss} = 0$ (see Fig.1). The control index $R_c$ is seen to reach unity around the resonance energies, giving complete control of $\sigma_{sup \to reac}$, in accord with considerations regarding control around resonances.[44] The reaction cross section can be tuned from 0 (destructive interference) to the sum $\sigma_{a \to reac} + \sigma_{b \to reac}$ by varying the superposition parameters $\eta$ and $\beta$. For example, the control landscape at two resonances energies ($E/E_{th} = 0.48$ and $0.59$) are illustrated in Fig. 2, showing large control ranging from 140 a.u. to 85069 a.u. (factor of 607) for the first resonance, and from 17 a.u. to 9264 a.u. (factor of 544) for the second. Note that the minima are not exactly zero due to the presence of small direct reactive scattering component.

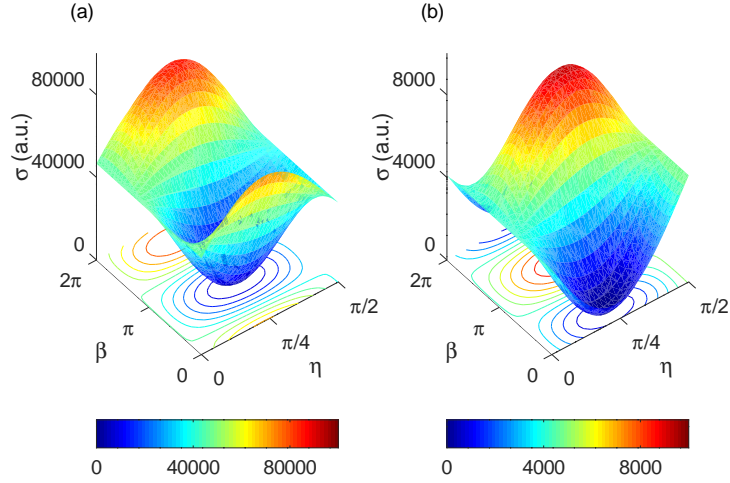

Figure 2: Control landscapes for the reactive cross sections at the resonance energies $E/E_{th}$=(a) 0.48 and (b) 0.59

Complete control around resonance energies results from the factorization of the S-matrix elements from the initial channel $|i\rangle$ to a product channel $|p\rangle$ into an initial state factor and a product state factor:

$$S_{i \to p} = \frac{\gamma_{ir}\gamma_{rp}}{(E - E_r) + i\Gamma^T/2}, \tag{6}$$

where $\gamma_{ir}$ and $\gamma_{rp}$ are the partial widths between the resonant state $r$ and the (ingoing/outgoing) scattering state in the corresponding open channel, and $\Gamma^T$ is the full width of the resonance.

As a result, the cross section from the superposition (1) also factorizes:

$$\sigma_{sup \to reac} = \frac{\pi}{k^2} \times \left|\cos\eta\gamma_{1r} + \sin\eta e^{i\beta}\gamma_{1r}\right|^2 \times \left(\sum_{p=3}^{N_f+2}\left|\frac{\gamma_{rp}}{(E-E_r)+i\Gamma^T/2}\right|^2\right). \quad (7)$$

This factorisation leads to control over complex formation, but not over the details of its dissociation. It allows complete suppression of complex formation by destructive interference. However, control only over complex formation implies non-selectivity of the product channels and hence no control of the branching ratio between the product channels.

On the other hand, the fact that control is over complex formation counters the deterious effect of proliferating product channels, as illustrated in Fig. 3 where the resonance ($s_{dir} = 0.0001$) and direct ($s_{dir} = 10$) reactions are compared. To obtain these results, we have generated 100 different samples (with $\rho = 5$, $x = 0.0001$, $w = 1$) for each $N_f$ and taken the average of the control index $R_c$ at the first resonance energy $E/E_{th} = 0.1$. These results have important implications for coherent control. Specifically, increasing the number of final channels generally suppresses control, as observed in Fig.3 for the direct reaction. This is the case because optimizing each final state may require a different set of control parameters, consequence of the change of the S-matrix elements. This variability decreases the value of $\sigma_{int}$ and then of the control index. In short, the global optimization is a complicated trade-off between the optimizations for each final states. By contrast, an isolated resonance allows for similar control landscape for each of the final states so that the control over a large number of final channels becomes possible.

The results in Fig. 3 motivate further study. Hence, we have examined the transition between the two opposite cases in Fig 3 by increasing the scaling factor $s_{dir}$ from 0.0001 to 0.1, fixing the number of final channels to 100 and keeping the same parameters (positions and widths) for the resonances. The decrease of the control index is illustrated, for the first resonance ($E/E_{th} = 0.1$), in Fig. 4. This result suggests that by measuring the control index, we could determine whether the reaction is driven by isolated resonances or whether the direct component is significant. Thus, the utility of exploring quantum scattering phenomena from initial superpositions extends beyond control, providing a new tools to probe scattering dynamics. Further work on this perspective is planned for the future.

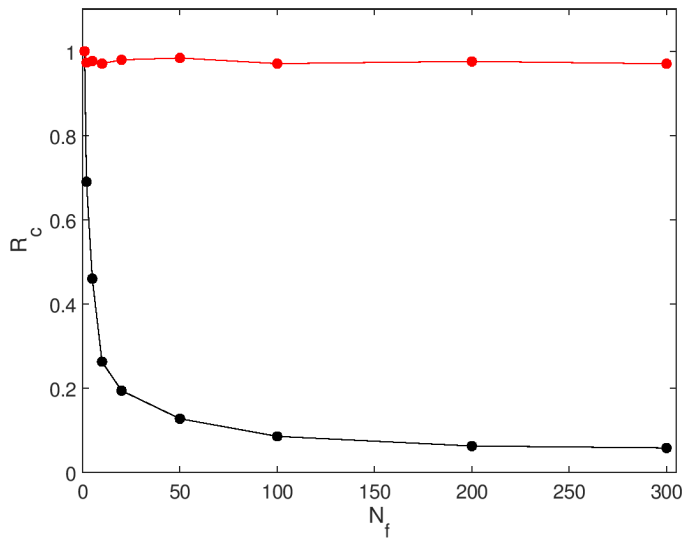

Figure 3: Dependence of control index around the resonance energy $E/E_{th} = 0.1$ averaged between 100 resonances samples on the number of product channels for direct scattering, $s_{dir} = 10$,(black) and resonance scattering, $s_{dir} = 10$ (red)

Finally, the regime of lossy resonances has been studied by increasing the value of $\Gamma_{loss}$, starting from the sample of Fig.1. The resonance parameters (resonance energies and widths) are kept at the same values and the direct contribution parameter is fixed at $s_{dir} = 0.0001$. The increase of $\Gamma_{loss}$ causes resonance broadening and overlap, leading to a reduction of the control index (see Fig. 5). The decrease of control is related to the ratio of $\Gamma_{loss}$ to the mean level distance $d$. For small values of $\Gamma_{loss}/d$, the resonances do not overlap and the results are the same as previously discussed. When $\Gamma_{loss}$ is of the same order of magnitude as $d$, a few resonances start to overlap and a reduction of the control index is observed. Finally, when $\Gamma_{loss}$ is larger than $d$ ($\Gamma_{loss}/d>2$), many resonances overlap and control is lost. Due to the overlap, the cross-section at a single energy no longer factorizes and is the sum of contributions of many broad resonances which would have been optimized with different optimal superposition parameters. This difference of optimal superposition parameters is the direct results of the random distribution of the partial widths and causes the decrease of $\sigma_{int}$ and then the loss of control. This is an alternate route to statistical scattering, distinct from the better-known energy-averaging over many isolated resonances,[27] discussed in the rest of the paper.

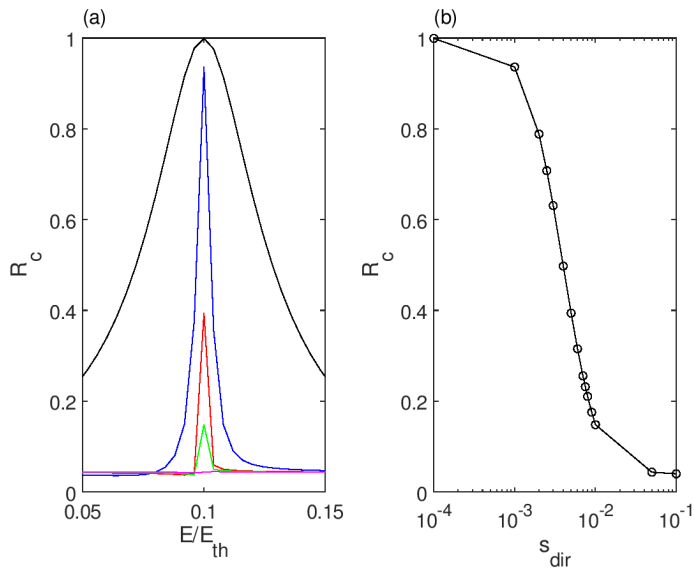

Figure 4: (a) Control index around the first resonance energy $E/E_{th} = 0.1$ for $s_{dir}$=0.0001 (black), 0.001(blue), 0.005 (red), 0.01 (green) and 0.05 (magenta) (b) Dependence of control index on the direct scattering scaling factor.

**Control of energy-averaged quantities (e.g. reaction rates).**– In some experiments, the measured quantity includes an energy average, either by definition (in the case of reaction rates) or by lack of resolution, e.g. where the Maxwell-Boltzmann distribution of the molecular gas is too broad to resolve the resonances. Here, we examine the impact of energy averaging by considering control of the reaction rate.

The thermal distribution shown in Fig. 1 (c) is spread over a large number of resonances. Then, in obtaining the rate, thermal averaging causes a significant loss of control (see Fig. 6), resulting in only 0.19 for the control index. This loss of control by thermal averaging comes from the difference of the optimal control parameters of the initial coherent superposition (1) between the resonances. These are determined by the partial widths $\gamma_{1r}$ and $\gamma_{2r}$: $\eta^r_{min} = \arctan(|\gamma_{1r}|/|\gamma_{2r}|)$ while $\beta^r_{min}$ is given by the difference of phases between $\gamma_{1r}$ and $\gamma_{2r}$. For example, the superposition parameters for the minimum values differ between the two resonances in Fig. 2: $\eta^r_{min} = 0.83$, $\beta^r_{min} = 2.36$ for the first resonance while $\eta^r_{min} = 0.74$, $\beta^r_{min} = 0.59$ for the second resonance. The value of $|\gamma_{1r_1}\gamma^*_{2r_1}E_1 e^{-E_1/E_{th}} + \gamma_{1r_2}\gamma^*_{2r_2}E_2 e^{-E_2/E_{th}}|$ depends on the difference $\beta^{r_1}_{min} - \beta^{r_2}_{min}$ and is small when $\beta^{r_1}_{min} - \beta^{r_2}_{min}$ is around $\pi$. The contributions of the two resonances then cancel one another in the thermal average giving a small interference contribution $\left|\int \sigma_{int} E e^{-E/E_{th}} dE\right|$ to the rate and to $R_{c,rate}$, where $\sigma_{int}$ is defined below Eq. (4). As a consequence, a given superposition can induce constructive interference at one resonance but destructive interference at another, as illustrated in the Fig. 7 (a). In other words, from the same initial superposition, the formation of the complex can be suppressed at one energy but enhanced at another energy. For multiple resonances, the small value of $\left|\int \sigma_{int} E e^{-E/E_{th}} dE\right|$ is connected to the random distribution of the partial widths and then of the optimal superposition parameters, illustrated on Fig. 7 (b) for the resonance sample of Fig 1.

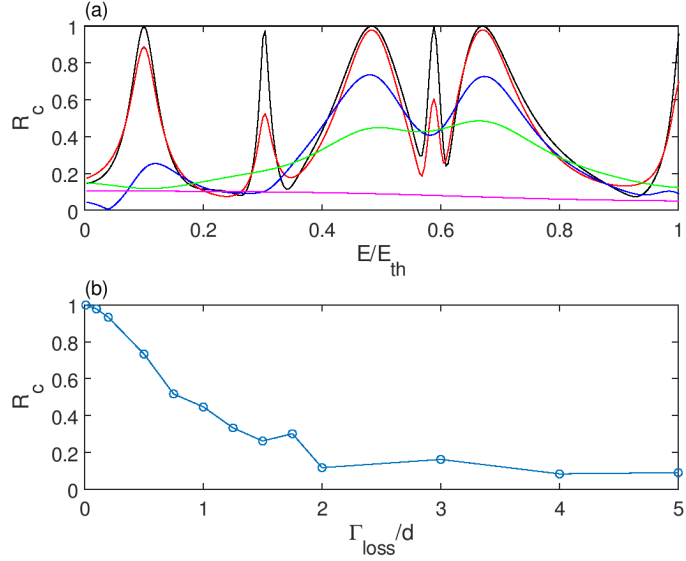

Figure 5: (a) Control index from the resonance sample in fig. 1 with different values of $\Gamma_{loss}/d=$ 0.01 (black), 0.1 (red), 0.5 (blue), 1(green) and 5 (magenta). (b) Dependence of control index on $\Gamma_{loss}/d$ around the resonance energy $E/E_{th} = 0.48$.

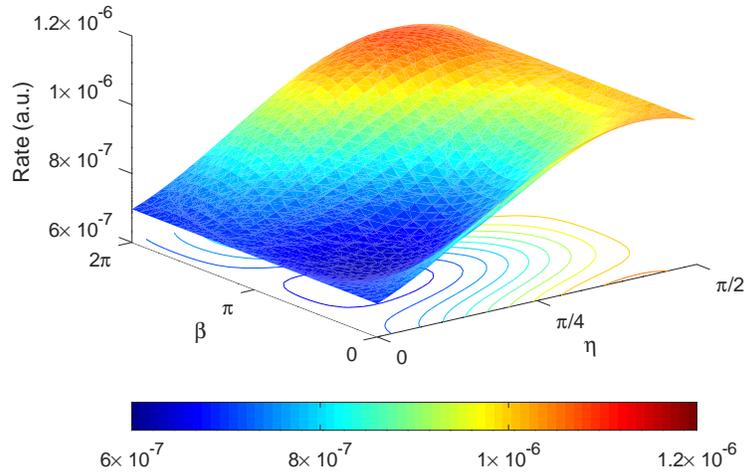

Figure 6: Control landscape for the reaction rate, obtained by thermal averaging of the cross section in Fig. 1

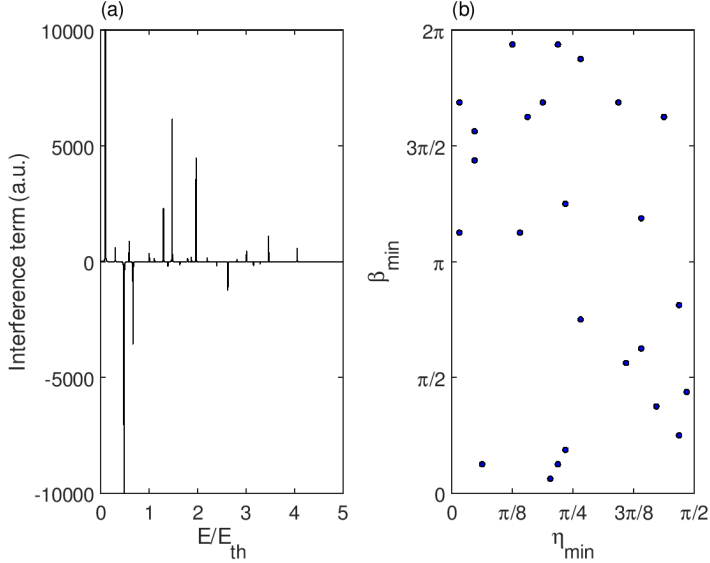

Figure 7: (a) Energy dependence of the interference term $2\sin\eta\cos\eta\ Re(\sigma_{int}\ e^{-i\beta})$ for $\eta = 0.83$ and $\beta = 2.36$. (b) Distribution of the optimal parameters ($\eta_{min}$, $\beta_{min}$) for different resonances.

The number of points in the Fig. 7 (b) is related to the resonance density $\rho$, leading to a $\rho$-dependence of the control index. To study the $\rho$-dependence, we generate 100 samples for each value of $\rho$ (with $x = 0.0001$, $s_{dir} = 0.0001$, and $w = 1$), from which we calculate the average values of $R_{c,rate}$ and its cumulative probability distribution $P(R_{c,rate}<\text{p})$, giving the probability that $R_{c,rate}$ is lower than p. The calculated cumulative probability distributions for different $\rho$ (see Fig. 8) are shifted to lower values of the control index $R_{c,rate}$, showing a reduced control. From this cumulative probability, we calculated the probability for having $R_{c,rate}$ between 0.5 and 1, setting $R_{c,rate} = 0.5$ as a threshold for a large control (see Fig. 9 (a)). For $\rho > 5$ $(E_{th})^{-1}$, the probability $P(0.5 < R_{c,rate} < 1)$ mostly vanishes (as shown in Fig. 9 (a)), illustrating that efficient control becomes unlikely. The complete vanishing of the control index $R_{c,rate}$ only happens at a larger $\rho$, as shown by the decrease of the average value in Fig. 9 (b), scaling as $1/\sqrt{\rho}$. Note that the origin of the loss of control for the rate due to thermal average over a large number of resonances is analogous to the loss of control over the direct scattering cross section due to proliferation of final channels (black line in Fig. 3). The reduction of the average value is also accompanied with that of the standard deviation. There is less and less variability between different samples as $\rho$ increases. Therefore, the validity of the statistical theory and the ability to control energy-averaged quantities are closely related.[45–48]

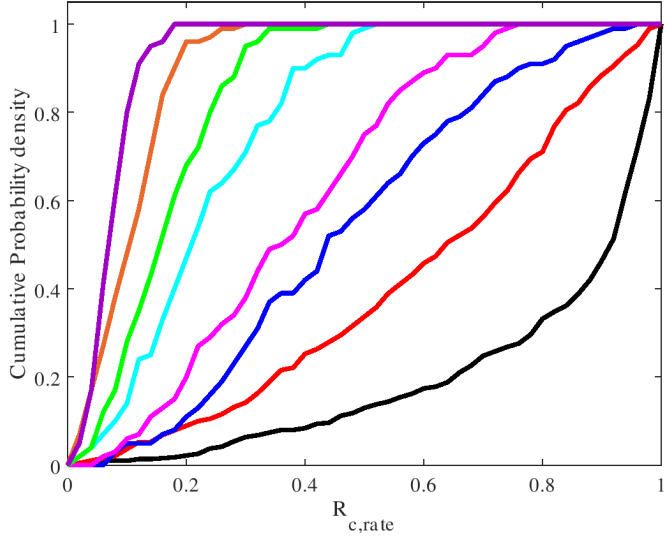

Figure 8: Cumulative probability distribution for $\rho$=0.2 (black), 0.5 (red), 1 (blue), 2 (magenta), 5 (cyan), 10 (green), 20 (orange) and 50 $(E_{th})^{-1}$ (purple).

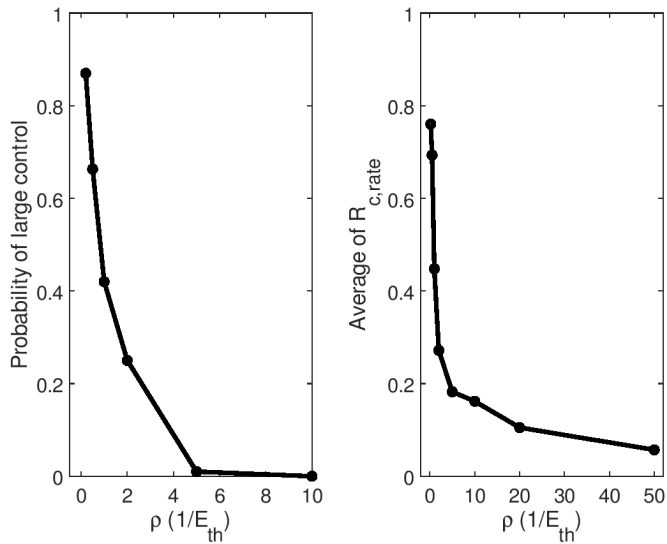

Figure 9: Decrease of the control index with $\rho$. (a) Probability of large control (see text for the definition) and (b) the average value of $R_{c,rate}$ plotted as a function of the resonance density $\rho$.

**Conclusions.**– Scattering resonances are seen to have two distinct roles with respect to control via quantum interference, i.e. via coherent control. If the system and experimental apparatus permit resolving isolated resonances then extensive control over the scattering cross section can be achieved at resonance energies. By contrast, in the case of the overlap of many broad resonances or in measurements that involve averages over resonances, such as those of scattering rates, control is suppressed as the system approaches the limit of statistical scattering.[22] These results imply that experiments allowing for reduced thermal averaging, such as in the optical collider[49–51] or the tomography of resonances,[52] can afford an improved platform for coherent control and provide deep insight into ultracold scattering insofar as averaging over collision energies can hide the underlying physics.

We note that magnetic scattering resonances have been observed in ultracold atom-molecule scattering, such as $^{23}$Na$^{40}$K + $^{40}$K,[38] and in ultracold NaLi+NaLi collisions.[53] Since the density of resonances is generally lower in atom-diatom scattering, these systems present desirable opportunities for coherent control of scattering observable at low temperatures. Computations on systems of this type are being planned in our laboratory as are computational studies of the control index $R_c$, which would afford insight into the relative contributions of direct scattering vs collision complex formation.

**Acknowledgements.**– This work was supported by the U.S. Air Force Office of Scientific Research (AFOSR) under Contract No. FA9550-19-1-0312.

# Supporting information for Coherent Control of ultracold molecular collisions: The role of resonances

Adrien Devolder[1], Timur V. Tscherbul[2], and Paul Brumer[1]

[1]*Chemical Physics Theory Group, Department of Chemistry, and Center for Quantum Information and Quantum Control, University of Toronto, Toronto, Ontario, M5S 3H6, Canada*

[2]*Department of Physics, University of Nevada, Reno, NV, 89557, USA*

E-mail:

# S1. Short-range K-matrix

The short-range K-matrix between an initial channel (i=1 or 2) and a product channel ($p = 3, 4, ...N_f + 2$) is constructed as the sum of two terms: the direct contribution and the resonance contributions, describing the reactions without or with formation of a complex, respectively:

$$K_{ip}^{sr} = K_{ip}^{dir} + \sum_{r=1}^{N_r} \frac{\gamma_{ir}\gamma_{rp}}{E - E_r + i\frac{\Gamma_{loss}}{2}}. \tag{1}$$

where $K_{ip}^{dir}$ and the resonance parameters ($E_r$ (resonance energy) and $\gamma_{ir}, \gamma_{rp}$ (partial width)) are randomly generated in our model.[1] The direct contribution is the product of a random Hermitian matrix and a scale factor $s_{dir}$, which determines the extent of the direct contribution. The number of resonances $N_r$ is determined by the resonance density $\rho$, given in units of $(E_{th})^{-1}$, where $E_{th} = k_B T$ is the thermal energy ($k_B$ is the Boltzmann constant and $T = 500$ nK is the temperature of the molecular ultracold gas). $\Gamma_{loss}$ is the loss rate to unobserved channels, e.g. after photoexcitation or collisions with a third molecule, and we consider the same value for every resonance. The quantities $\gamma_{ir}$ and $\gamma_{rp}$ are the partial widths between the resonant state $r$ and the (ingoing/outgoing) scattering state in the corresponding open channel. Their moduli are normally distributed:[1,2]

$$P(|\gamma|) = \sqrt{\frac{\pi\rho}{2x}} \exp\left(\frac{-\pi^2\rho}{2x}|\gamma|\right) \tag{2}$$

The variance of the distributions $\frac{x}{\rho\pi^2}$ is parameterized by $x$, related to the overlap of the resonances (in absence of loss to unobserved channels), and by $\rho$. A small value for $x$ corresponds to the isolated resonances case while a large value corresponds to the overlapping resonances case. Moreover, $x$ can be related to the loss parameter in the Multichannel Quantum Defect Theory.[3] Regarding the phase of the partial widths, they are generated following an uniform distribution between 0 and $2\pi$. The total width of a resonances $\Gamma_r^T = \Gamma_{loss} + 2\pi(|\gamma_{1r}|^2 + |\gamma_{2r}|^2) + 2\pi \sum_{p=3}^{N_f+2} |\gamma_{rp}|^2$ is the sum of the loss rate $\Gamma_{loss}$ to the unobserved channels and the width due to the coupling to the open channels explicitly included in the model. Finally, the generation of the resonance energies $E_r$ is given by parametrization of the nearest-neighbor energy spacing between the resonances, which follows the distribution:[1,2]

$$P(s) = (1+w)A(w)s^w \exp(-(A(w)s)^{1+w}), \qquad (3)$$

where $A(w) = \Gamma\left(\frac{w+2}{w+1}\right)$, $\Gamma(.)$ is the Gamma function, $s = |E_r - E_{r-1}|/d$ is the spacing between adjacent levels scaled by the mean level distance $d = 1/\rho$ and $w$ is the Brody parameter, determining the chaotic character of the scattering. The result is a Poisson distribution for $w = 0$ and a Wigner distribution for $w = 1$ (chaotic system). The energy of the first resonance is fixed at $E_1 = d/2$. The resonance energies $E_r$ are scaled with respect to the thermal energy $E_{th}$, as are the other quantities of the model ($\rho$, $\Gamma_{loss}$, $\gamma_{ir}$ and $\gamma_{rp}$).

## S2. Long-range Scattering matrix

The relationship between the scattering matrix and the short-range K-matrix (Eq. (1)) is given by MQDT[2–4] as:

$$\boldsymbol{R} = \boldsymbol{C}^{-1}\left((\boldsymbol{K}^{sr})^{-1} - \tan\boldsymbol{\lambda}\right)^{-1}\boldsymbol{C}^{-1}, \qquad (4)$$

$$\boldsymbol{S} = e^{i\boldsymbol{\xi}}[1 + i\boldsymbol{R}][1 - i\boldsymbol{R}]^{-1}e^{i\boldsymbol{\xi}}, \qquad (5)$$

where $\boldsymbol{\xi}$, $\boldsymbol{C}$ and $\tan\boldsymbol{\lambda}$ are diagonal matrices of quantum defect parameters. The two initial channels are near threshold, giving the diagonal elements:[4]

$$C_{ii} = \sqrt{k\bar{a}_i(s_i - 1)^2}, \qquad (6)$$

$$\tan\lambda_{ii} = 1 - s_i, \qquad (7)$$

$$\xi_{ii} = \arctan(-k\bar{a}_i s_i), \qquad (8)$$

where $k$ is the relative momenta, $\bar{a}_i$ is the mean scattering length and $s_i = \frac{a_i}{\bar{a}_i}$ is the relative scattering length. In our simulations, we use values from KRb-KRb scattering studied experimentally at $T$=500 nK: $\bar{a}_i = 117.59\ a_0$ and $s_i = 1.2$.[3]

On the other hand, the reactive channels are far from threshold, giving the diagonal elements:[4] $C_{ii} = 1$, $\tan\lambda_{ii} = 0$ and $\xi_{ii} = 0$.